\documentclass[structabstract]{aa}  
\usepackage{subfigure}
\usepackage{graphicx}
\usepackage{txfonts}
\usepackage{natbib}
\bibpunct{(}{)}{;}{a}{}{,}

\begin{document}
   \title{A \textit{Herschel}\thanks{\textit{Herschel} is an ESA space observatory with science instruments provided by European-led Principal Investigator consortia and with important participation from NASA.} Resolved Far-Infrared Dust Ring around HD~207129}

%   \subtitle{}

   \author{Jonathan~P.~Marshall\inst{1}
          \and T.~L\"ohne\inst{2}
          \and B.~Montesinos\inst{3}
          \and A.V.~Krivov\inst{2}
          \and C.~Eiroa\inst{1}
          \and O.~Absil\inst{4}\fnmsep\thanks{FNRS Postdoctoral Researcher}
          \and G.~Bryden\inst{5}
          \and J.~Maldonado\inst{1}
          \and A.~Mora\inst{6,1}
          \and J.~Sanz-Forcada\inst{3}
          \and D.~Ardila\inst{7}
          \and J.-Ch.~Augereau\inst{8}
          \and A.~Bayo\inst{9}
          \and C.~del~Burgo\inst{10}
          \and W.~Danchi\inst{11}
          \and S.~Ertel\inst{12}
          \and D.~Fedele\inst{1,13,14}
          \and M.~Fridlund\inst{15}
          \and J.~Lebreton\inst{8}
          \and B.M.~Gonz\'alez-Garc\'ia\inst{16}
          \and R.~Liseau\inst{17}
          \and G.~Meeus\inst{1}
          \and S.~M\"uller\inst{2}
          \and G.L.~Pilbratt\inst{15}
          \and A.~Roberge\inst{11}
          \and K.~Stapelfeldt\inst{5}
          \and P.~Th\'ebault\inst{18}
          \and G.J.~White\inst{19,20}
          \and S.~Wolf\inst{12}
          }

   \institute{Departmento F\'isica Te\'orica, Facultad de Ciencias, Universidad Aut\'onoma de Madrid, Cantoblanco, 28049, Madrid, Spain\\
         \email{jonathan.marshall@uam.es}
         \and Friedrich-Schiller-Universit\"at Jena, Astrophysikalisches Institut und Universit{\"a}tssternwarte, 07743, Jena, Germany
         \and Departmento de Astrof\'{\i}sica, Centro de Astrobiolog\'{\i}a (CAB, CSIC-INTA), ESAC Campus, P.O. Box 78, 28691\\
              Villanueva de la Ca\~nada, Madrid, Spain
         \and Institut d'Astrophysique et de G{\'e}ophysique, Universit{\'e} de Li{\`e}ge, 17 All{\'e}e du Six Ao{\^u}t, B-4000 Sart Tilman, Belgium
         \and Jet Propulsion Laboratory, California Institute of Technology, Pasadena, CA 91109, USA
         \and ESA-ESAC Gaia SOC. P.O. Box 78 E-28691 Villanueva de la Ca{\~n}ada, Madrid, Spain
         \and NASA Herschel Science Center, California Institute of Technology, 1200 E. California Blvd., Pasadena, CA 91125, USA
         \and UJF-Grenoble 1 / CNRS-INSU, Institut de Plan\'etologie et d'Astrophysique de Grenoble (IPAG) UMR 5274, Grenoble, F-38041, France
         \and European Space Observatory, Alonso de Cordova 3107, Vitacura, Casilla 19001, Santiago 19, Chile
         \and UNINOVA-CA3, Campus da Caparica, Quinta da Torre, Monte de Caparica, 2825-149 Caparica, Portugal
         \and NASA Goddard Space Flight Center, Exoplanets and Stellar Astrophysics, Code 667, Greenbelt, MD 20771.USA
         \and Christian-Albrechts-Universit\"at zu Kiel, Institut f\"ur Theoretische Physik und Astrophysik, Leibnizstr. 15, 24098 Kiel, Germany
         \and Max-Planck Institut f\"ur Astronomie, K\"onigstuhl 17, 69117 Heidelberg, Germany
         \and John Hopkins University, Dept. of Physics and Astronomy, 3701 San Martin drive, Baltimore, MD 21210, USA
         \and ESA Astrophysics \& Fundamental Physics Missions Division, ESTEC/SRE-SA,  Keplerlaan 1, NL-2201 AZ Noordwijk,\\
              The Netherlands
         \and INSA at ESAC, E-28691 Villanueva de la Ca{\~n}ada, Madrid, Spain
         \and Onsala Space Observatory, Chalmers University of Technology, Se-439 92 Onsala, Sweden
         \and LESIA, Observatoire de Paris, 92195 Meudon Principal Cedex, Paris, France
         \and Department of Physics and Astrophysics, Open University, Walton Hall, Milton Keynes MK7 6AA, UK
         \and Rutherford Appleton Laboratory, Chilton OX11 0QX, UK
             }

   \date{Received ---; accepted ---}

   \abstract{Dusty debris discs around main sequence stars are thought to be the result of continuous collisional grinding of planetesimals in the system. The majority of these systems are unresolved and analysis of the dust properties is limited by the lack of information regarding the dust location.}{The Herschel DUNES key program is observing 133 nearby, Sun-like stars ($<$ 20~pc, FGK spectral type) in a volume limited survey to constrain the absolute incidence of cold dust around these stars by detection of far infrared excess emission at flux levels comparable to the Edgeworth-Kuiper belt (EKB).}{We have observed the Sun-like star HD~207129 with \textit{Herschel} PACS and SPIRE. In all three PACS bands we resolve a ring-like structure consistent with scattered light observations. Using $\alpha$ Bo\"otis as a reference point spread function (PSF), we deconvolved the images, clearly resolving the inner gap in the disc at both 70 and 100~$\mu$m.}{We have resolved the dust-producing planetesimal belt of a debris disc at 100~$\mu$m for the first time. We measure the radial profile and fractional luminosity of the disc, and compare the values to those of discs around stars of similar age and/or spectral type, placing this disc in context of other resolved discs observed by \textit{Herschel}/DUNES.}{} 
% 5 {} token are mandatory
 

   \maketitle

%
%________________________________________________________________

\section{Introduction}

Debris discs are composed of dust grains continuously produced by the collisional grinding of larger unseen planetesimals. This is inferred from the short lifetime of the dust grains compared to the age of the star \citep{bp93} and suggest that the star around which they are observed has undergone a planetesimal formation process. In almost all cases, dust production is consistent with steady state attrition of the dust parent bodies \citep{ace_paper}. The vast majority of debris discs are unresolved and we can only use modelling of the disc spectral energy distribution (SED) to determine the spatial location of the dust and the physical properties of the constituent dust grains. In the absence of independent information constraining the spatial location of the dust, which introduces degeneracies into the fitting of the dust grain properties, e.g. between grain size and radial location, a standard dust composition is assumed. In the few examples of resolved debris discs, structures that imply the presence of a planetary mass body to maintain and sculpt the disc are frequently observed, e.g. warps, asymmetries and blobs \citep{golmi06,kalas05,sheret04}. Planets may also be responsible for the inner cavities seen in resolved discs and inferred from the SEDs of unresolved ones \citep{wyatt08}. The resolved disc structure can therefore be used as an indirect probe for exoplanets around such stars in regions of orbital radius/planetary mass parameter space that are otherwise inaccessible to traditional search methods (e.g. radial velocity, transits) and remain a challenge to direct imaging techniques.

HD~207129 (HIP~107649) was identified as having a debris disc by \textit{IRAS} \citep{ref_iras} and followed up by both \textit{ISO} and \textit{Spitzer} \citep{iso_obs,trilling08}. Extended emission from the disc has been seen in both scattered light (\textit{HST}) and in thermal infrared emission \citep[\textit{Spitzer} MIPS70,][]{spt_obs}. The disc excess was also detected at 160~$\mu$m by both \cite{ex_160} and \cite{spt_obs}, though with very different values (155~mJy cf 250~mJy), illustrating the need for \textit{Herschel} PACS observations to constrain the disc SED. Additionally, the system has been identified as a promising candidate for exoplanet searches, due to its proximity (16~pc), age (1--3~Gyr) and the presence of an inner gap in the disc \citep{iso_obs,exo_tgt}. 

In this paper we present \textit{Herschel} \citep{herschel_ref} PACS \citep{pacs_ref} and SPIRE \citep{spire_ref} observations of HD~207129. The large aperture \textit{Herschel} telescope provides arcsecond resolution allowing detailed imaging of the debris disc in this system. In addition, greater precision and denser coverage of the disc SED can be obtained. Altogether, this allows better constraints to be placed on the disc's SED and physical extent compared to previous observations, thereby allowing additional refinement of our models of this solar system analogue.

\section{Observations and data reduction}

HD~207129 was observed as part of the DUNES \citep[DUst around NEarby Stars;][Rodmann et al., in prep.]{eiroa10} volume limited survey of nearby ($d <$ 20~pc) Sun-like (FGK) stars. PACS scan map observations of the star were taken with both 70/160 and 100/160 channel combinations. Each scan map consisted of 10 legs of 3$'$ length, with a 4$''$ separation between legs, at the medium slew speed (20$''$ per second). The target was observed at two position angles (70 and 110$^{\circ}$) in both wavelength combinations. SPIRE small map mode observations were also carried out on a separate Observation Day (OD) covering a region $\sim$~4$'$ around HD~207129 at the nominal slew speed (30$''$/s). Using five repetitions of the scan map observations reduced the expected noise level to close to that of the expected extragalactic contribution (7--9~mJy at 250--500~$\mu$m) and increased the coverage of the central region of the map allowing pixel sizes smaller than the standard values to be used in the image reconstruction process (i.e. image scales of 4$''$, 6$''$ and 8$''$ per pixel at 250, 350 and 500~$\mu$m, limited by the appearance of gaps in the image coverage near the centre of the map), which was useful for looking at extended structure in the source brightness profile. A summary of the observations is presented in Table \ref{obs_log}.

\subsection{\textit{Herschel} photometry}

PACS data reduction was carried out in HIPE 4.2 (the latest available public release\footnote{see: $\rm{http://herschel.esac.esa.int/HIPE{\_}download.shtml}$}), starting from the level 0 products using the standard reduction script. The separate scans at the two position angles of each channel pair were mosaiced to produce a final image at each wavelength. Image scales for the final mosaics were 1$''$ per pixel for the blue (70/100~$\mu$m) images and 2$''$ per pixel for the red (160~$\mu$m) image. A high-pass filter was used to remove large scale background emission from the images, with filter widths of 15 and 25$''$ in the blue and red channels, respectively.  A central region of 30$''$ radius in the images was masked from the high pass filter process to prevent the removal of any faint extended structure near to the source. SPIRE data reduction was also carried out in HIPE 4.2, again starting from the level 0 data using the standard script and processing options. The SPIRE maps were created using the naive scan mapper alogrithm. The pixel scales for the SPIRE 250, 350 and 500~$\mu$m images from which photometry was taken were 6$''$, 10$''$ and 14$''$.  

\begin{table}
\caption{Summary of \textit{Herschel} observations of HD~207129.}             
\label{obs_log}      
\centering          
\begin{tabular}{lcccr} 
\hline\hline
Instrument & Observation ID  & OD & Wavelengths & Duration  \\
           &                 &    & [$\mu$m]    & [s]       \\
\hline
PACS       & 1342193163/64  & 322 & 70/160      & 276.0  \\                             
PACS       & 1342193165/66  & 322 & 100/160     & 2250.0 \\                                         
SPIRE      & 1342209300     & 544 & 250/350/500 & 721.0  \\
\hline                  
\end{tabular}
\end{table}

PACS fluxes were measured using aperture photometry carried out using the IDL APER routine. The aperture radius and sky annulus dimensions were 20$''$ and 30--40$''$, respectively. SPIRE fluxes were measured using an aperture of radius 30$''$ (due to the presence of several nearby sub-mm bright background objects), whilst the sky noise values were taken from \cite{nguyen10}.

\begin{table}
\caption{Summary of \textit{Herschel} and ancillary infrared/sub-mm photometry used in disc SED fitting. Uncertainties 1-$\sigma$ and driven by the sky noise. \textit{Herschel} calibration uncertainties are 10\% for PACS and 15\% for SPIRE.}  
\label{ir_phot}      
\centering          
\begin{tabular}{lll} 
\hline\hline
 Wavelength & F$_{obs}$  & Instrument \\
 $[ \mu m]$ & [mJy]      &            \\
 \hline
  9 & 1237$~\pm~$17& \textit{AKARI}/IRC PSC, \cite{irc_psc}\\
 18 & 263$~\pm~$31& \textit{AKARI}/IRC PSC, \cite{irc_psc}\\
 24 & 155$~\pm~$5.3& \textit{Spitzer}/MIPS, \cite{trilling08}. \\
 32 & 111$~\pm~$5.1& \textit{Spitzer}/IRS \\
 70 & 278$~\pm~$11& \textit{Spitzer}/MIPS,  \cite{trilling08}\\
 160 & 250$~\pm~$40& \textit{Spitzer}/MIPS, \cite{spt_obs} \\
 160 & 158$~\pm~$20& \textit{Spitzer}/MIPS, \cite{ex_160}\\
 70  & 284$~\pm~$1.5 & \textit{Herschel}/PACS\\
 100 & 311$~\pm~$1.1 & \textit{Herschel}/PACS\\
 160 & 211$~\pm~$1.5 & \textit{Herschel}/PACS\\
 250 & 113$~\pm~$18 & \textit{Herschel}/SPIRE\\
 350 & 44.3$~\pm~$9 & \textit{Herschel}/SPIRE\\
 500 & 25.9$~\pm~$8 & \textit{Herschel}/SPIRE\\
 870 & 5$~\pm~$3& APEX/LABOCA, \cite{apx_obs}\\
 
\hline                  
\end{tabular}
\end{table}

\subsection{Stellar parameters}

HD~207129 (HIP~107649) is a nearby \citep[$d\!=\!16~\pm~0.2$~pc,][]{ref_dist} star with a reported spectral type of G2V \citep[from the \textit{Hipparcos} catalogue,][]{hip_cat} or G0V \citep{gray06}. The bolometric luminosity has been estimated from the absolute magnitude and bolometric correction using measurements by \cite{flower96}. Our adopted values for the effective temperature, gravity and metallicity are derived from the mean of spectroscopic measurements from \cite{santos04}, \cite{vf05} and \cite{sousa08}. Our own estimate of the rotational velocity, from high resolution spectra, is $\upsilon$~sin~$i =~3.71~\pm~$1.81~kms$^{-1}$ (Maldonado et al., in prep.), consistent with estimates of \cite{groot96} and \cite{torres06}. Assuming the stellar inclination is that of the disc ($i~=~$60$~\pm~$5$^{\circ}$, see Sect. 3.2) and the stellar radius is 1.07~R$_{\odot}$ (estimated from the bolometric luminosity and effective temperature), the rotational period of the star is $P~=~$12.6~days. The star is non-active, with measurements of the activity index, $\log~R'_{\rm{HK}}$, of $-~4.8$ \cite{henry96} and $-~5.02$ \cite{gray06}, and a \textit{ROSAT} X-ray luminosity log~$L_{\rm{X}}/L_{\rm{bol}}~=~-5.63$.

The stellar mass estimated from the radius and gravity is 1.15~$M_{\odot}$. From Padova evolutionary tracks \citep{girardi02}, a mass of 1.0--1.1~$M_{\odot}$ and age of $\sim$~3.2~Gyr are obtained. This age is consistent with the activity index of \cite{henry96}, using the calibration from \cite{mh08}. Conversely, the age derived from the rotational period using the same calibration is $\sim$~1.6~Gyr, which is consistent with the X-ray luminosity age of 1.8~Gyr using the relationship of \cite{garces10}, or the results of \cite{giardino08} (their Fig. 8). Other age estimates range from 0.6 \citep{song03} to 6.0~Gyr \citep{lachaume99,vf05,holmberg09}. We measure a LiI 6708~$\AA$ equivalent width of 35.5$~\pm~$3.3~m$\AA$ (Maldonado et al., in prep.); this value, in conjunction with the estimate derived from the rotation period, points to an age greater than 600~Myr, implying the stellar age lies between these extremes. We therefore adopt an age in the range 1.5--3.2~Gyr as appropriate for HD~207129 \citep[see Figs 6 and 9,][]{maldonado10}. The stellar parameters and observational properties are summarised in Table \ref{star_params}. 

The stellar photosphere contribution to the total flux was computed, using the stellar parameters, from a synthetic stellar atmosphere model interpolated from the PHOENIX/GAIA grid \citep{brott05}. Optical and near infrared photometry including Stromgren $uvby$, Tycho-2 $BV$ and $JHK$ from \cite{aumann91} constrain the stellar component of the SED, which has been scaled to the \textit{Spitzer} IRS spectrum following the method of \cite{bertone04}.

\begin{table}
\caption{Physical properties of HD 207129.}
\label{star_params}
\begin{tabular}{ll}
\hline\hline
Parameter   &  Value  \\
\hline
Distance  & 16$~\pm~$0.2~pc \\
Spectral type and luminosity class & G2V, G0V  \\
$V$, $B-V$ & 5.57, 0.60~mag\\
Absolute magnitude $M_V$, Bolometric Correction & 4.55, --0.06\\
Bolometric luminosity, $L_*$ & 1.258~L$_\odot$ \\ 
Effective temperature & 5912~K \\
Surface gravity, $\log g$ & 4.44  \\
Radius, $R_*$ & 1.07~$R_\odot$ \\
Metallicity, $[\rm{Fe/H}]$ & --0.01 \\
Rotational velocity, $\upsilon \sin i$ & 3.71~kms$^{-1}$ \\
Rotation period, $P$ &  $\sim$ 12.6~days  \\
Activity, $\log R'_{\rm{HK}}$ & --4.80, --5.02  \\
X-Ray luminosity, $\log~L_{\rm{X}}/L_*$ & --5.63\\
Mass, $M_*$ &  1.0--1.15~$M_\odot$ \\
Age         & 1.5--3.2~Gyr\\
\hline
\end{tabular}
\end{table}

\section{Results}

\subsection{Disc spectral energy distribution}

For the purposes of disc SED modelling, the PACS photometry was supplemented by a broad range of infrared and sub-millimetre observations including \textit{AKARI} IRC all-sky survey 9/18~$\mu$m \citep{irc_psc}, \textit{Spitzer} IRS spectrum and MIPS 24 and 70~$\mu$m photometry \citep{trilling08}, \textit{Spitzer} 160~$\mu$m \citep{spt_obs}, APEX/LABOCA 870~$\mu$m \citep{apx_obs} fluxes and upper limits from 450/850~$\mu$m  JCMT/SCUBA \citep{sheret04} and 1.2~mm SEST \citep{schuetz05} observations. The \textit{Herschel} photometry and all complementary data are summarised in Table \ref{ir_phot} (the upper limits are not quoted in the table because they were not used in the SED fitting process).

\begin{figure}[htb]
\centering
\includegraphics[width=0.5\textwidth]{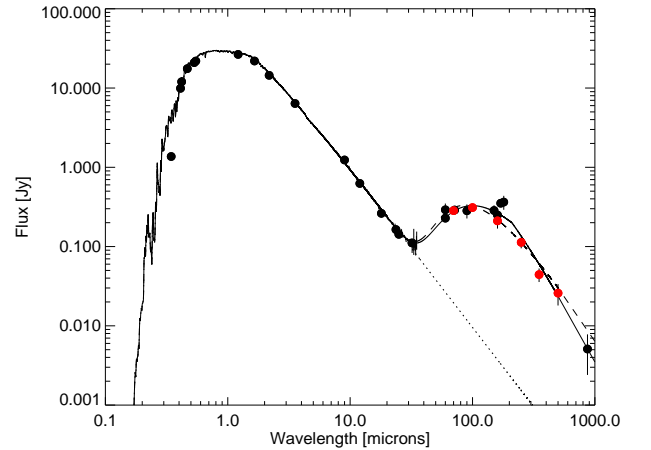}
\caption{HD~207129 spectral energy distribution. Red data points are the PACS and SPIRE photometry, black data points are the ancillary data, including the \textit{Spitzer} IRS spectrum. Error bars are 1~$\sigma$ and include both calibration uncertainty (10\% for 70/100~$\mu$m, 20\% for 160~$\mu$m) and sky noise. Some error bars are smaller than the data point symbol. For wavelengths longer than $\sim~30~\mu$m the solid line represents a disc model with a $\lambda_{0} = 210~\mu$m  and $\beta = 1$, the dashed line represents a  disc model with a $\lambda_{0} = 63~\mu$m  and $\beta = 2.7$ and the dotted line represent the stellar photosphere.}
\label{final_sed}
\end{figure}

The disc SED, with the \textit{Herschel} fluxes and ancillary photometry can be seen in Fig \ref{final_sed}. We have fitted the excess emission from HD~207129 with a standard black body model, modified beyond 210~$\mu$m by a factor of $\beta$~=~1 \citep[see Eqn 6, ][]{wyatt08}, from which estimates of the disc fractional luminosity and disc orbital radius were derived. The model was fitted to the infrared (both \textit{Herschel} and ancillary) photometry through least squares minimisation weighted by the observational uncertainties. The best fit temperature $T_{\rm{disc}} = 50$~K and fractional luminosity $L_{\rm{IR}}/L_{\rm{Bol}} = 8.3\times10^{-5}$ were obtained with a reduced $\chi^{2}$ of 1.34. Using the derived stellar parameters and the typical disc temperature, we calculate a dust orbital radius of 34~AU for black body emission, clearly much smaller than the resolved disc which would imply that a disc orbital radius larger than that derived from the black body modelling is required.

\begin{figure}[ht]
\centering
\includegraphics[width=0.5\textwidth]{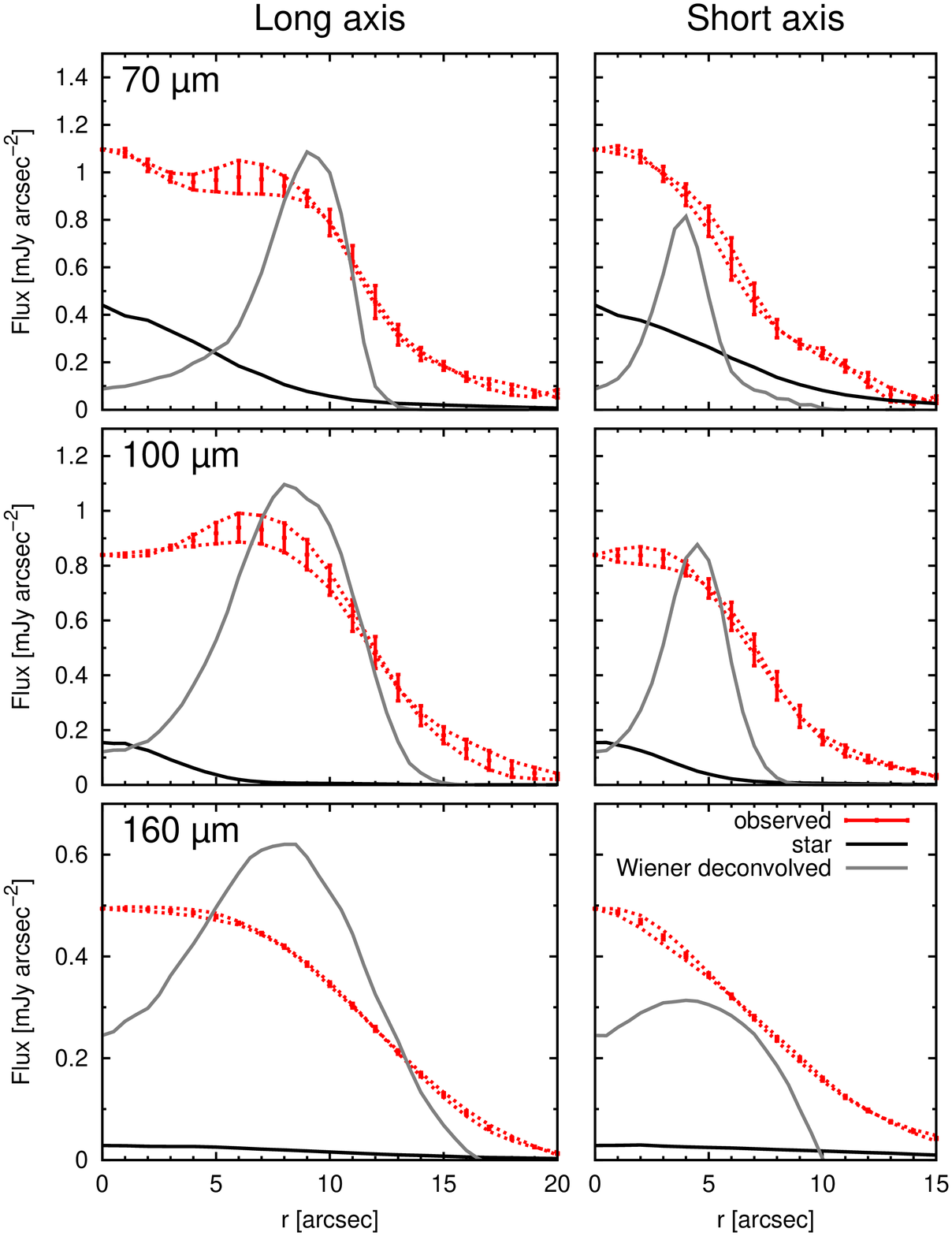}
\caption{Radial profiles of HD~207129 along the semi-major (left) and semi-minor (right) axes of the disc. The red dotted lines are the measured profile either side of the disc centre along each axis; the solid red error bars mark the mean position at each point and the associated uncertainty. The grey line is the deconvolved disc profile and the black line is the subtracted stellar profile.}
\label{deconv_graph}
\end{figure}

\subsection{Disc images}

The disc of HD~207129 is clearly resolved in the PACS mosaics at all three wavelengths and is extended in the two shorter SPIRE wavelength maps. The optical position of HD~207129 is (in the epoch of the \textit{Herschel} images) 21$^{h}$48$^{m}$16.05$^{s}$ -47$^{\circ}$18$'$14.47$''$ \citep[using proper motions from the re-reduction of \textit{Hipparcos},][]{ref_dist}. In the PACS 70~$\mu$m image there is a peak in the disc brightness $\sim$~2$''$ NE from this position, consistent with the optical position within the \textit{Herschel} pointing uncertainty, with a flux value consistent with the predicted stellar photosphere contribution. If this peak were the star, that would imply that the disc was asymmetric. In the deconvolution presented here, we have assumed that the star is at the centre of the disc, consistent with the symmetric scattered light disc observed in the \textit{HST} results \citep{spt_obs}, and that the peak observed in the 70~$\mu$m image is a dust blob.

The disc size and position angle were measured by fitting a rotated ellipse to the source brightness profile contour of 3 times the sky noise value in the PACS images, whilst the disc inclination is measured from the ratio of the semi-minor to semi-major axes. A summary of these properties is in Table \ref{disc_orient}. The disc extent is the same in both the 70 and 100~$\mu$m images, though we would naively expect that the apparent disc size would increase due to the larger beam size and greater contribution from colder dust emission. We attribute the similar disc sizes to a combination of the similar beam sizes at the two wavelengths (5.7$''$ and 6.7$''$ FWHM at 70 and 100~$\mu$m, respectively) which are both small compared to the extent of the disc. In both the 70~$\mu$m and 100~$\mu$m image, there are two peaks in the source brightness profile, with the NW peak $\sim$~10\% brighter than the SE one (at 100~$\mu$m). This structure is consistent with the underlying disc being a symmetric ring-like structure. Additionally at 70~$\mu$m, there is a third peak close to the centre of the disc, which is NE ($\sim$~2$''$) of the disc centre and stellar position, which may be the result of inhomogenities in the disc structure. The disc source brightness profile at 160~$\mu$m (and in the SPIRE images) is smooth, showing no internal structure.

\begin{table}
\caption{Measurement of the disc extent before (left -- the HWHM of a Gaussian fitted to the source brightness profile) and after (right -- the radial extent of annulus peak brightness) deconvolution, and the inclination and position angle of the disc at all three wavelengths.}
\label{disc_orient}
\centering
\begin{tabular}{lrrrrrr}
\hline\hline
          & \multicolumn{6}{c}{Wavelength}\\
          & \multicolumn{2}{c}{70~$\mu$m}& \multicolumn{2}{c}{100~$\mu$m}& \multicolumn{2}{c}{160~$\mu$m}\\
\hline                    
Semi-major axis [$''$]       & 14.5  & 9.0 & 14.0 & 9.0 & 17.0 & 8.5 \\  
Semi-major axis [AU]         & 232   & 144 & 224  & 144 & 272  & 136 \\ 
Semi-minor axis [$''$]       &  8.5  & 4.0 & 8.5  & 4.5 &   12.0 & 4.0 \\
Semi-minor axis [AU]         & 136   & 64 & 136  & 72 &   192.0 & 64 \\ 
Position angle$^{a}$ [$^{\circ}$] & 120 &  & 122 &  & 120  & \\
Inclination$^{b}$ [$^{\circ}$] &  54   & 64 &  53  &  60 &   45 & 62 \\
\hline
\multicolumn{7}{l}{$^{a}$The disc position angles are the same in both original and}\\
\multicolumn{7}{l}{$~$$~$deconvolved images.}\\
\multicolumn{7}{l}{$^{b}$The disc minor axis is not resolved at 160~$\mu$m in the original images.}\\
\end{tabular}
\end{table}

In order to measure the true extent of the disc, the images were deconvolved from the instrument PSF, a technique which has previously proved successful \citep{liseau10}. An observation of $\alpha$ Bo\"otis was used as the PSF model for the deconvolution, rotated to match the roll angle of the telescope at the time of observing HD~207129.

Deconvolution was a two step process; the stellar photosphere contribution was removed from each image by subtraction of a PSF with a peak scaled to the predicted photospheric flux level in that image and centred on the stellar position determined from measurement of the 70~$\mu$m isophotes. The optical and isophote derived positions are in good agreement and the small offset between them does not impact upon the findings of this analysis. 

\begin{figure*}[h]
\centering
\subfigure{\includegraphics[width=0.3\textwidth]{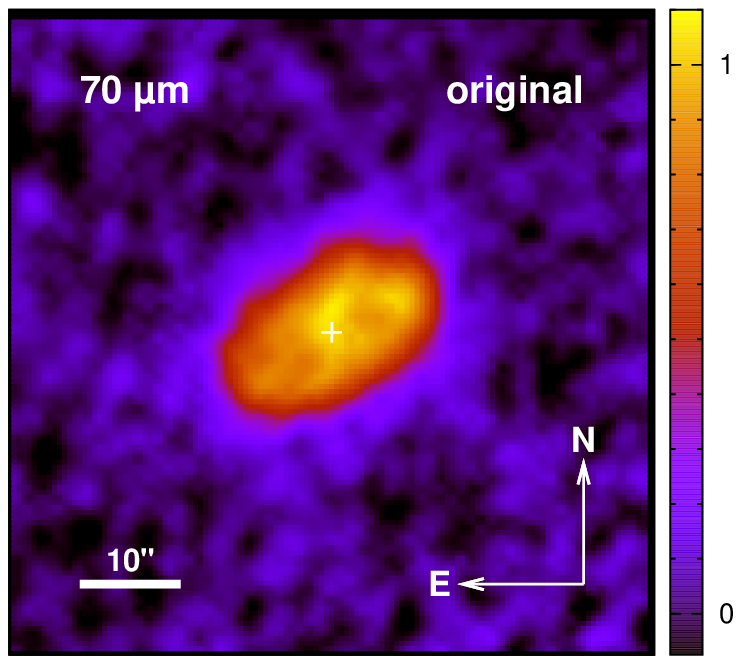}}
\subfigure{\includegraphics[width=0.3\textwidth]{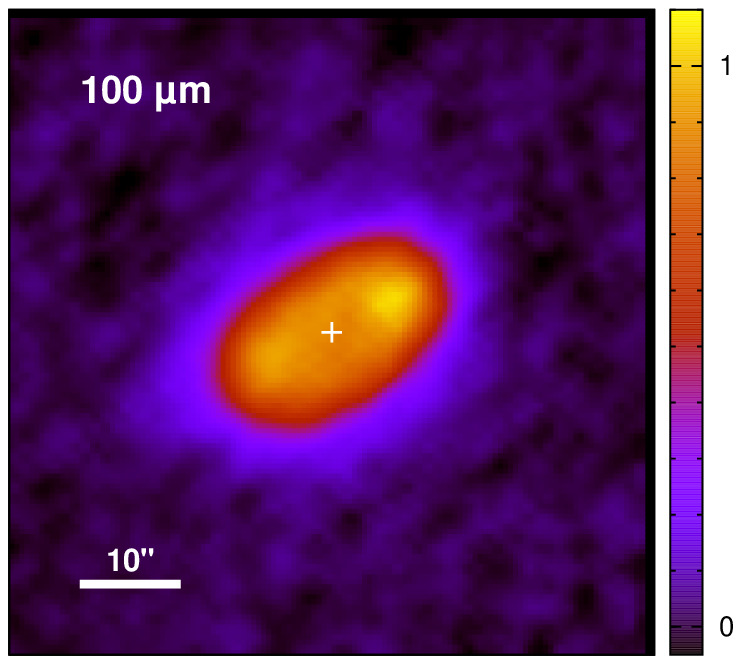}}
\subfigure{\includegraphics[width=0.3\textwidth]{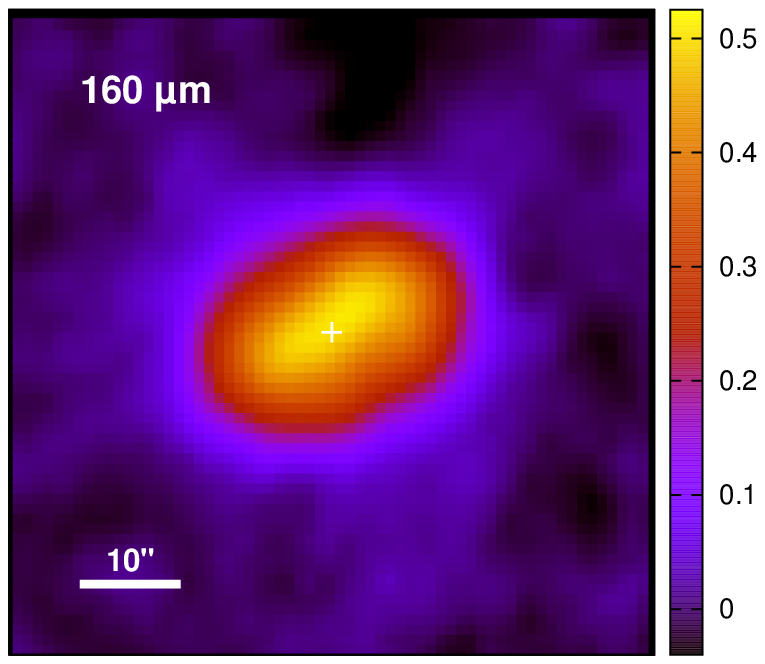}}
\subfigure{\includegraphics[width=0.3\textwidth]{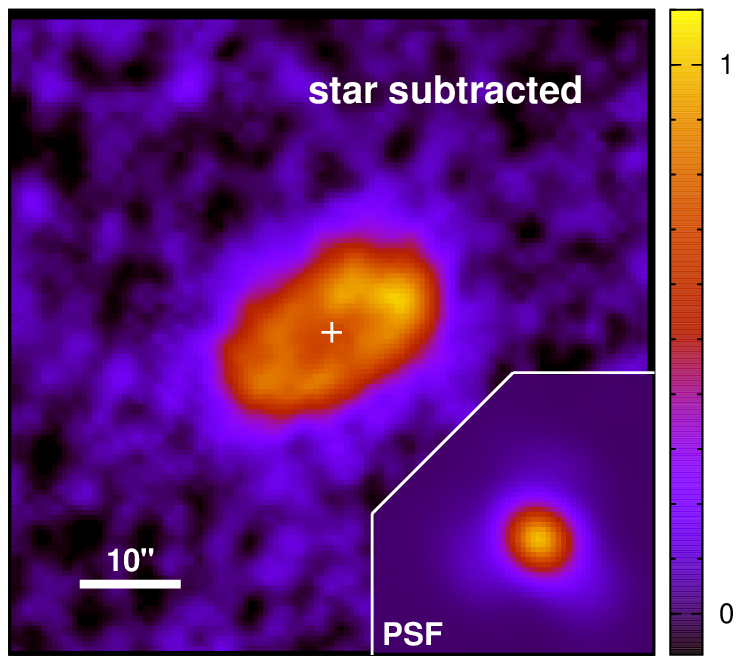}}
\subfigure{\includegraphics[width=0.3\textwidth]{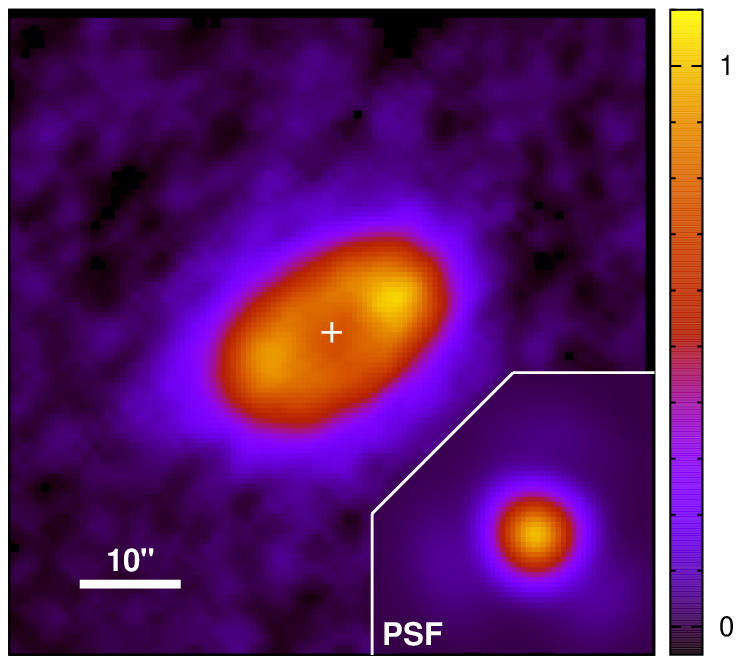}}
\subfigure{\includegraphics[width=0.3\textwidth]{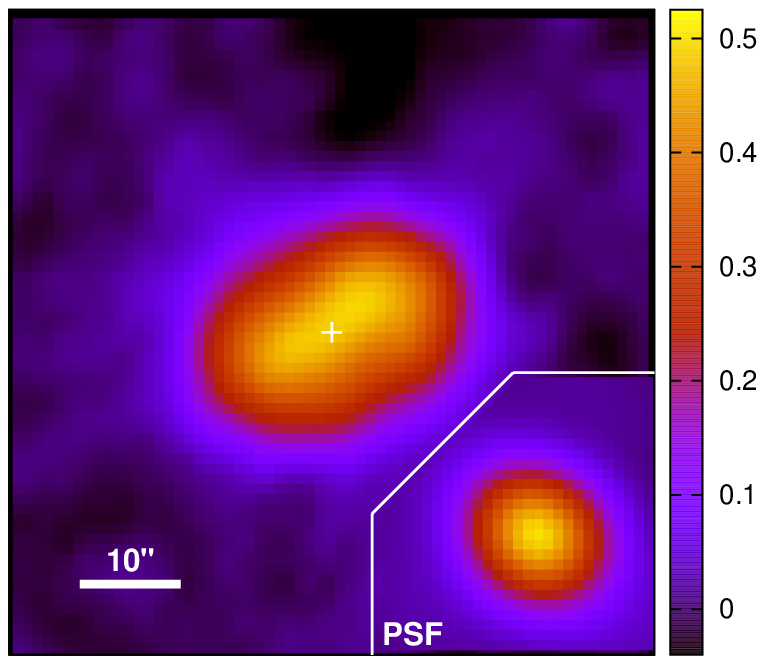}}
\subfigure{\includegraphics[width=0.3\textwidth]{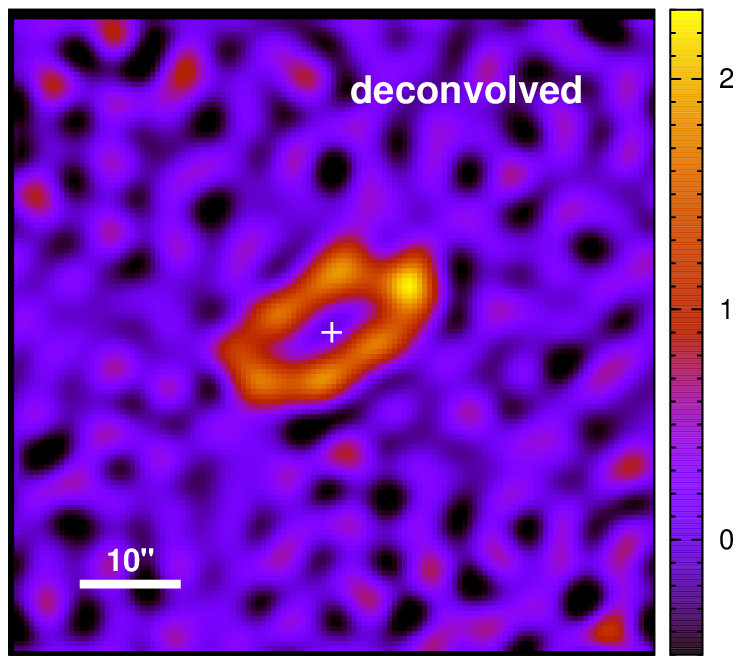}}
\subfigure{\includegraphics[width=0.3\textwidth]{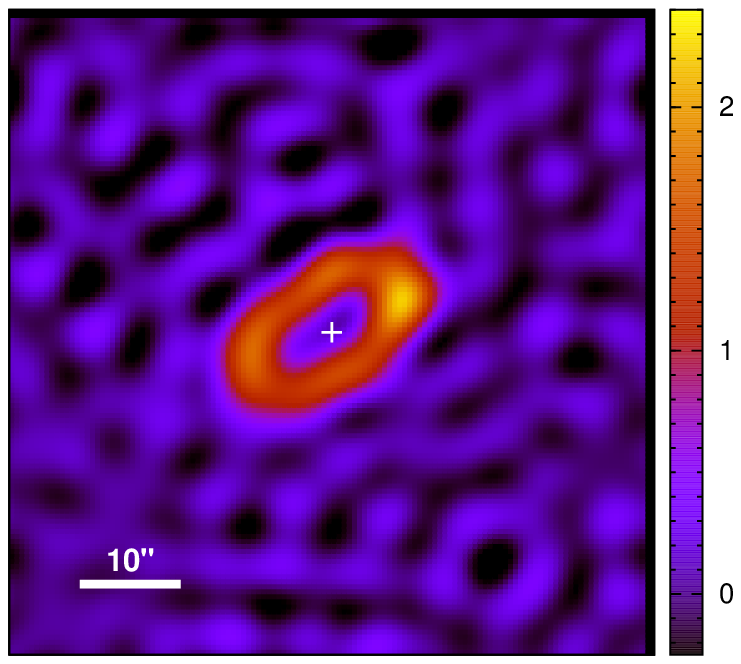}}
\subfigure{\includegraphics[width=0.3\textwidth]{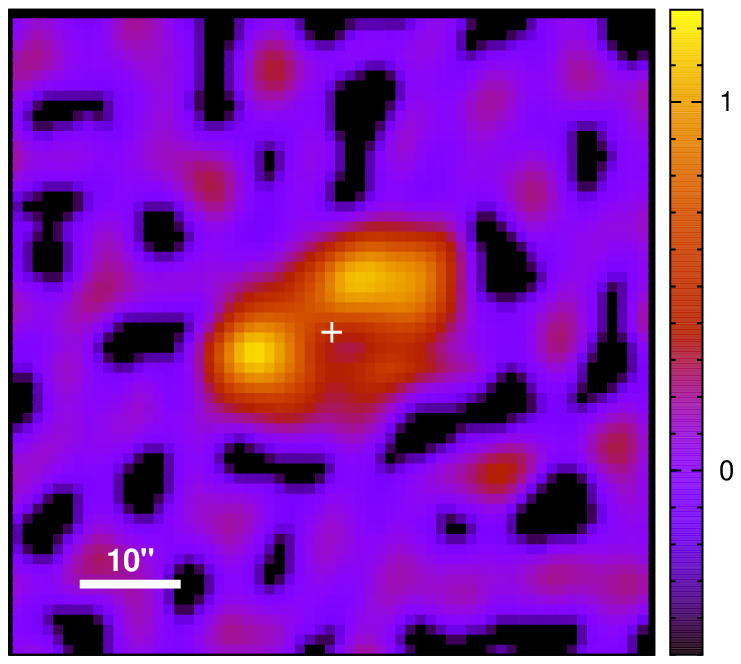}} 
\caption{Original image (top), post stellar subtraction (middle) and Wiener deconvolved image (bottom) in the three PACS wavelengths. The ring structure is clearly visible in both the 70 and 100~$\mu$m images. Image orientation is North up, East left. A distance scale bar has been provided, 10$''$ is equivalent to 160~AU. The colour scale bar is in units of mJy/arcsec$^{2}$ . The stellar position, derived from isophotes in the 70 and 100~$\mu$m observations, is marked as a white cross in each image.}
\label{deconv_img}
\end{figure*}

After star subtraction, the image was deconvolved using three separate methods (modified Wiener, Richardson-Lucy and van Cittert) to check the suitability of the individual methods and the repeatability of any structure observed in the deconvolved images. It was found that all three methods clearly produced a ring structure in the deconvolved 70 and 100~$\mu$m images, though with varying noise patterns. Deconvolution of the 160~$\mu$m image via modified Wiener and van Cittert methods resulted in images with the disc structure as a pair of blobs either side of the stellar position. Using the Richardson-Lucy method the disc structure at 160~$\mu$m was recovered as a broken ring surrounding the stellar position with a clear central gap. We do not interpret the clumpy blobs in the deconvolved disc images as representing real structure in the debris disc. The result of the image deconvolution using the modified Wiener method can be seen in Fig. \ref{deconv_img}.

The deconvolved radial extent of the disc was measured from the position of the peak brightness of the observed disc annulus along its major and minor axes. In the deconvolved 100~$\mu$m image, the measured radial extent was 9$''$~$~\pm~$2$''$ (144$~\pm~$32~AU, see Fig \ref{deconv_graph}), with similar values from the 70 and 160~$\mu$m images.  We measure the position angle of the disc to be 120$~\pm~$5$^{\circ}$ through fitting via a least squares minimisation algorithm. From the ratio of the semi-major and semi-minor axes, the inclination to be 54$~\pm~$5$^{\circ}$ (60$~\pm~$5$^{\circ}$ in the deconvolved images). These results are consistent with previous \textit{HST} measurements \citep{spt_obs}. We see no shift in the positional angle of the disc toward longer wavelengths.

\subsection{Discussion}

We have directly resolved the structure of a ring-like debris disc for the first time in the far infrared. The disc radial extent of $\sim$~140~AU is comparable to that of Fomalhaut \citep{kalas05} or HD~107146 \citep{ardila04}. 

The disc radius derived from the standard black body model, modified beyond 210~$\mu$m by a factor of $\beta$~=~1, was 34~AU. This is a large underestimate of the true extent, in direct conflict with the resolved images. It serves to illustrate the dangers of modelling debris discs using only the SED and/or black body thermal emission models, which neither assume, nor tell you anything about, the dust grain optical properties \citep{wyatt08}. Using a disc of dust grains that emit as modified black bodies and 10~$\mu$m dust grains, with a break shortward of the SED peak, the disc radius derived from the SED matches the observed images but fails to reproduce the sub-millimetre slope. The observed dust optical and thermal emission properties cannot be reconciled via Mie theory, either. A disc composed of small ($<~1~\mu$m) dust grains would, according to Mie theory, reproduce the symmetric scattered light image, though the dust would then be too warm to reproduce the SED \citep{spt_obs}.

We measure a dust fractional luminosity of 8.3$\times$10$^{-5}$, around six times greater than Vega \citep[$1.5\times10^{-5}$,][]{habing01} and half that of q$^{1}$ Eri \citep[$1.5\times10^{-4}$,][]{liseau08}. The value we measure is consistent with the fractional luminosity/age relation from \cite{decin03}. The disc extent, $\sim$~140~AU, is larger than other DUNES resolved discs, e.g. q$^{1}$ Eri \citep[85~AU,][]{liseau10} or $\zeta^{2}$ Ret \citep[70--120~AU,][]{eiroa10}, but the derived black body temperature is also large, $\sim$~50~K, cf 60~K for the q$^{1}$ Eri disc which is around an F star and 30--40~K for the $\zeta^{2}$ Ret disc which has a smaller extent around a star of similar spectral type.

A complete analysis of the disc SED and physical structure, in which both classical power-law particle size distribution models and a self-consistent collisional model are fitted to the available photometric and imaging data will be presented by L\"ohne et al., 2011 (in prep.).

\section{Conclusions}

We have presented \textit{Herschel} PACS and SPIRE observations of HD~207129, the first to directly resolve the ring-like structure of a debris disc in both major and minor axes at far infrared wavelengths. The disc extent (140~$\pm$32~AU), inclination (51~$\pm$~5$^{\circ}$) and typical (black body) temperature ($\sim$~50~K), derived purely from the observations assuming no particular grain model, are similar to previous measurements. Compared to other discs around stars of similar spectral type and age, the disc of HD~207129 is both larger and warmer (for the observed size), though it is not completely atypical in either respect, but situated at the margins of the known range of disc morphologies. A simple analysis based on a black body model has been proved unrealistic. A more detailed self-consistent study is left for future work (L\"ohne et al, in prep.). 

\bibliographystyle{aa}

\bibliography{16673}

\begin{acknowledgements}
We would like to thank the staff at the Herschel Science Centre, discussions with whom added immeasurably to the quality of the data reduction. C.~Eiroa, J.~Maldonado,  J.P.~Marshall, and B.~Montesinos are partly supported by Spanish grant AYA 2008/01727. T.~L\"ohne, S.~M\"uller and A.~Krivov acknowledge support by the DFG, projects Lo 1715/1-1 and Kr 2164/9-1. J.-Ch.~Augereau, J.~Lebreton and P.~Th\'ebault are supported by a CNES-PNP grant.

\end{acknowledgements}

\end{document}